\documentclass[aps,prl,twocolumn,superscriptaddress,10pt,showpacs]{revtex4-1}
\pdfoutput=1

\usepackage[utf8x]{inputenc}
\usepackage{graphicx}
\usepackage{amsmath,amssymb}
\usepackage[svgnames]{xcolor}
\usepackage{datetime}
\usepackage{color}
\usepackage[
pdftitle={Exploring competing density order in the ionic Hubbard model with ultracold fermions},    % title
pdfauthor={Michael Messer, Remi Desbuquois, Thomas Uehlinger, Gregor Jotzu, Sebastian Huber, Daniel Greif, Tilman Esslinger},     % author
pdfsubject={Ionic Hubbard},   % subject of the document
pdfcreator={Michael Messer, Remi Desbuquois, Thomas Uehlinger, Gregor Jotzu, Sebastian Huber, Daniel Greif, Tilman Esslinger},   % creator of the document
pdfkeywords={Ionic Hubbard} {Optical Lattice} {Ultracold Fermions}, % list of keywords
colorlinks=True,linkcolor=DarkBlue,citecolor=DarkBlue,urlcolor=DarkBlue,
pdfstartview=FitH,bookmarks=False,pdfpagemode=UseNone
]{hyperref}

\begin{document}

\title{Exploring competing density order in the ionic Hubbard model \\ with ultracold fermions}

\author{Michael Messer}
\affiliation{Institute for Quantum Electronics, ETH Zurich, 8093 Zurich, Switzerland}

\author{Rémi Desbuquois}
\affiliation{Institute for Quantum Electronics, ETH Zurich, 8093 Zurich, Switzerland}

\author{Thomas Uehlinger}
\affiliation{Institute for Quantum Electronics, ETH Zurich, 8093 Zurich, Switzerland}

\author{Gregor~Jotzu}
\affiliation{Institute for Quantum Electronics, ETH Zurich, 8093 Zurich, Switzerland}

\author{Sebastian Huber}
\affiliation{Institute for Theoretical Physics, ETH Zurich, 8093 Zurich, Switzerland}

\author{Daniel Greif}
\affiliation{Institute for Quantum Electronics, ETH Zurich, 8093 Zurich, Switzerland}

\author{Tilman Esslinger}
\affiliation{Institute for Quantum Electronics, ETH Zurich, 8093 Zurich, Switzerland}

%\date{\today}

\begin{abstract}
We realize and study the ionic Hubbard model using an interacting two-component gas of fermionic atoms loaded into an optical lattice. 
The bipartite lattice has honeycomb geometry with a staggered energy-offset that explicitly breaks the inversion symmetry. 
Distinct density-ordered phases are identified using noise correlation measurements of the atomic momentum distribution. 
For weak interactions the geometry induces a charge density wave. 
For strong repulsive interactions we detect a strong suppression of doubly occupied sites, as expected for a Mott insulating state, and the externally broken inversion symmetry is not visible anymore in the density distribution. 
The local density distributions in different configurations are characterized by measuring the number of doubly occupied lattice sites as a function of interaction and energy-offset. 
We further probe the excitations of the system using direction dependent modulation spectroscopy and discover a complex spectrum, which we compare with a theoretical model.
\end{abstract}


\maketitle

%P1:
Changes in the fundamental properties of interacting many-body systems are often determined by the competition between different energy scales, which may induce phase transitions. 
A particularly intriguing situation arises when the geometry of a system sets an energy scale that competes with the scale given by the interaction of its constituents. 
The importance of geometry is apparent in reduced dimensions which influences the interacting many-body system in its evolution from one phase to another \cite{Giamarchi2003}.  
A tractable approach to generic questions is provided by the ionic Hubbard model, which captures key aspects of the physics of a competing geometry and interactions in the charge sector. 
The Hamiltonian has a staggered energy-offset on a bipartite lattice, such that geometry supports a band insulating charge density wave (CDW). 
Conversely, strong repulsive on-site interactions favour a Mott insulating state (MI) at half-filling, which does not reflect the broken symmetry of the underlying lattice. 
The model was introduced in the context of charge-transfer organic salts \cite{Hubbard1981, Nagaosa1986} and has been proposed to explain strong electron correlations in ferroelectric perovskite materials \cite{Egami1993}.
Ultracold atoms in optical lattices are an excellent platform for studying competing energy scales, as they allow for tuning various parameters and the geometry of the Hamiltonian \cite{Lewenstein2012, Greiner2002, Sebby-Strabley2006, Joerdens2008, Schneider2008, Sias2008, Ma2011, Soltan-Panahi2011, Struck2011, Tarruell2012, Meinert2013, DiLiberto2014, Murmann2015}. 
Here we explore the ionic Hubbard Model using ultracold fermions loaded into a tunable optical honeycomb potential.

\begin{figure}[bt]
    \includegraphics{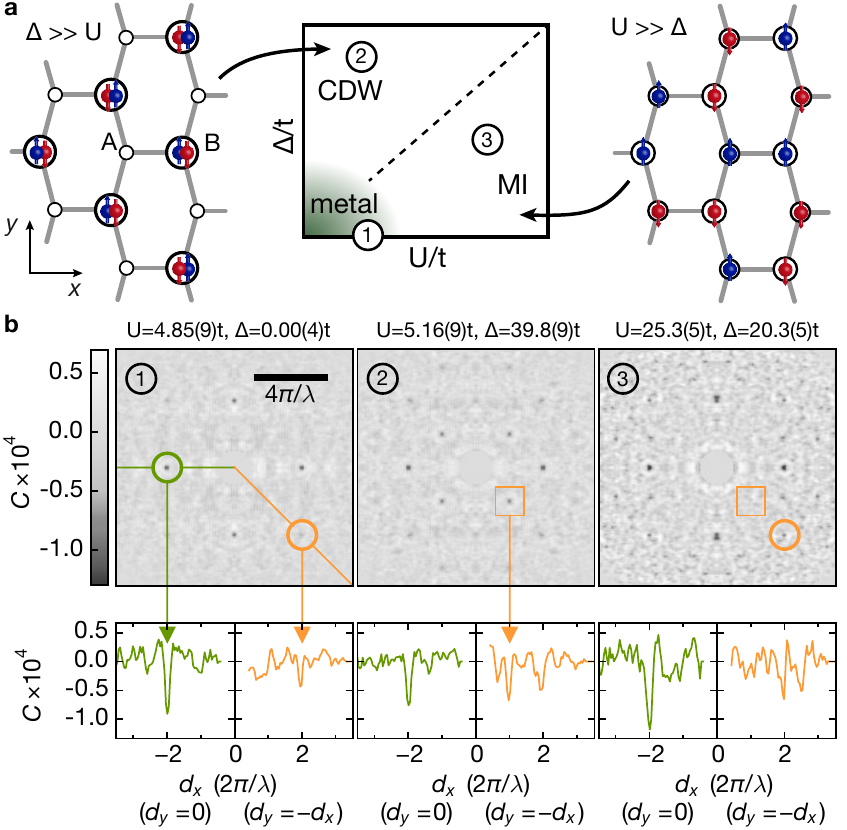}
    \caption{Noise correlations 
    {\bf (a)} Schematic view of the ionic Hubbard model on a honeycomb lattice at half-filling. Circles denote lattice sites {\bf A} and {\bf B}, where larger circles indicate lower potential energy. 
    The phase diagram exhibits two limiting cases: For $\Delta \gg U ,t$ a CDW ordered state is expected with two fermions of opposite spin (red, blue) on lattice sites {\bf B}, and empty sites {\bf A}. 
    In the other limit ($U \gg \Delta ,t$) a MI with one fermion on each lattice site should appear. 
    {\bf (b)} Measured noise correlation pictures obtained from absorption images of the atomic momentum distribution. 
    Comparing panel 1 with panel 2, additional correlations appear due to broken inversion symmetry in the CDW ordered phase.
    When introducing strong interactions, these correlations are not observed anymore (panel 3), and the inversion symmetry of the density distribution does no longer reflect the broken inversion symmetry of the lattice potential.   
    Below each panel horizontal and diagonal cuts of the noise correlation image are shown.
    For the three different ratios of $\Delta$ and $U$, between 165 and 201 measurements were taken each. 
    We show the average of $C(d_x,d_y)$ and $C(d_x,-d_y)$, which reflects the symmetry of the system.
	}\label{fig1}
\end{figure}

%P2
The ionic Hubbard model has been studied theoretically in 1D chains \cite{Fabrizio1999, Wilkens2001, Kampf2003, Manmana2004, Batista2004, Torio2006} and on the 2D square lattice \cite{Kancharla2007, Paris2007, Bouadim2007, Hoang2010}. 
More recently, these studies have been extended to a honeycomb lattice, motivated by possible connections to superconductivity in layered nitrides \cite{Watanabe2013} and strongly correlated topological phases \cite{Prychynenko2014}. 
We consider the ionic Hubbard model on a honeycomb lattice:
\begin{equation} 
\hat{H} = -t
\sum_{\langle ij \rangle,\sigma}\hat{c}_{i\sigma}^{\dagger}\hat{c}_{j\sigma} + U
\sum_{i} \hat{n}_{i\uparrow} \hat{n}_{i\downarrow} + 
\Delta \sum_{i \in \bf A,\sigma} \hat{n}_{i\sigma}, 
\label{hamiltonian}
\end{equation}
where $\hat{c}_{i\sigma}^{\dagger}$ and $\hat{c}_{i\sigma}$ are the creation and  annihilation operators of one fermion with spin $\sigma= \,\uparrow,\downarrow$ on site $i$ and $\hat{n}_{i\sigma}=\hat{c}_{i\sigma}^{\dagger}\hat{c}_{i\sigma}$.
The system is characterized by three energies: the kinetic energy denoted by the tunnelling amplitude $t$ and summed over nearest neighbours $\langle ij \rangle$, the on-site interaction $U$ and the staggered energy-offset between sites of the {\bf A} and {\bf B} sub-lattice $\Delta$, with $\Delta > 0$.
In addition there is a harmonic confinement in all three directions. 
All parameters of the Hamiltonian are computed using Wannier functions \cite{Uehlinger2013}.

%P3:
The interplay between the interaction energy $U$, the energy-offset $\Delta$ and tunnelling $t$ leads to quantum phases which differ by their density ordering. 
The two limiting cases can be qualitatively understood in the atomic limit at half-filling.  
For $U \gg \Delta$ the system is described by a MI state.
For a large energy-offset $\Delta \gg U$, we expect a band insulator with staggered density and two fermions on lattice site {\bf B} \cite{Kancharla2007}.
The resulting CDW pattern reflects the broken inversion symmetry of the underlying geometry. 
We can characterize the transition by an order parameter {$N_{\bf A} - N_{\bf B}$}, which is zero in the MI state or when $\Delta = 0$, with $N_{\bf A(B)}$ the total number of atoms on sub-lattice {\bf A(B)}.
Fig. \ref{fig1}a provides a schematic view of the different scenarios.

%P4:
In order to realize the ionic Hubbard model we create a quantum degenerate cloud of $^{40}\, \mathrm{K}$ as described in previous work \cite{Uehlinger2013} and detailed in \cite{supplementary}. 
We prepare a balanced fermionic spin mixture with total atom numbers between $1.5 \times 10^5$ and $2.0 \times 10^5$, with 10\% systematic uncertainty.
A $m_{F}=-9/2,-5/2$ ($m_{F}=-9/2,-7/2$) mixture with temperatures of 16(2)\% (13(2)\%) of the Fermi-temperature, is then loaded into a three-dimensional optical lattice within $200 \, \mathrm{ms}$. 
Using interfering laser beams at a wavelength $\lambda=1064 \, \mathrm{nm}$ we create a honeycomb potential in the xy-plane, which is replicated along the z-axis \cite{Tarruell2012, Uehlinger2013}.
All tunnelling bonds are set to $t/h=174(12)\, \mathrm{Hz}$. 
The tunable lattice allows us to independently adjust the energy-offset $\Delta = [0.00(4),41(1)]t$ between the {\bf A} and {\bf B} sub-lattice \cite{supplementary}.
Depending on the desired interaction strength we either use the Feshbach resonance of the $m_{F}=-9/2,-7/2$ mixture or the $m_{F}=-9/2,-5/2$ mixture.

%P5:
We probe the spatial periodicity of the density distribution in the interacting many-body state by measuring correlations in the momentum distribution obtained after time-of-flight expansion and absorption imaging \cite{Altman2004,Folling2005,Rom2006, Spielman2007, Guarrera2008, Simon2011}.  
After preparing the system in a shallow honeycomb lattice with a given $U$ and $\Delta$, we rapidly convert the lattice geometry to a deep simple cubic lattice.
This ensures that we probe correlations of the underlying density order rather than a specific lattice structure. 
The atoms are released from the lattice and left to expand ballistically for $10\, \mathrm{ms}$.
We then measure the density distribution, which is proportional to the momentum distribution of the initial state $n(\mathbf{q})$. 
From this, we compute the correlator of the fluctuations of the momentum distribution \cite{supplementary_nc_detection, Altman2004,Folling2005,Rom2006, Spielman2007, Guarrera2008, Simon2011},
\begin{equation} 
C(\mathbf{d}) = \frac{\int \langle n(\mathbf{q_{0}}-\mathbf{d}/2)\cdot n(\mathbf{q_{0}}+\mathbf{d}/2)\rangle \mathrm{d} \mathbf{q_{0}}} {\int \langle n(\mathbf{q_{0}}-\mathbf{d}/2) \rangle \langle n(\mathbf{q_{0}}+\mathbf{d}/2) \rangle \mathrm{d} \mathbf{q_{0}}} -1, 
\label{correlator}
\end{equation}
where the $\langle \rangle$ brackets denote the statistical averaging over absorption images taken under the same experimental conditions.

%P6:
Owing to the fermionic nature of the particles, this quantity exhibits minima when $\mathbf{d}=\mathbf{m} 2\pi/\lambda$, with $\mathbf{m}$ a vector of integers \cite{supplementary}. 
This is illustrated by the anti-correlations of a repulsively interacting, metallic state with $U=4.85(9)t$ and $\Delta=0.00(4)t$, shown in Fig. \ref{fig1}b, left panel. 
There, the spatial periodicity of the atomic density follows the structure of the lattice potential, and minima in the correlator are observed for $\mathbf{m}=(0,\pm 2)$ and $\mathbf{m}=(\pm 2,0)$. 
For $\Delta=39.8(9)t$, additional minima are observed at $\mathbf{m}=(\pm 1,\pm 1)$, see Fig. \ref{fig1}b, central panel. 
For a simple cubic lattice potential of periodicity $\lambda/2$, the amplitude of these minima is given by \cite{supplementary}
\begin{equation}
C\left(\pm\frac{2\pi}{\lambda},\pm\frac{2\pi}{\lambda}\right) \propto \frac{(N_{\bf A}-N_{\bf B})^2}{(N_{\bf A}+N_{\bf B})^2}.
\label{sublat_order}
\end{equation} 
Thus, the observation of additional minima confirms the presence of CDW-ordering with $N_{\bf A} \neq N_{\bf B}$. 
Finally, for $\Delta=20.3(5)t$ and $U=25.3(5)t$, these additional minima are not observed any more (see Fig. \ref{fig1}b, right panel), signalling that with repulsive on-site interactions, the density distribution does not reflect the externally broken inversion symmetry. 
In this case the interactions suppress the CDW-order, despite the presence of a large $\Delta$.

%P7:
Based on these measurements we expect the local distribution of atoms on each lattice site to depend on the exact values of $U$ and $\Delta$.
We measure the fraction of atoms on doubly occupied sites $D$ using interaction-dependent rf-spectroscopy \cite{Joerdens2008}.
The number of doubly occupied sites compared to the number of singly occupied sites is directly related to the nature of the insulating states \cite{Scarola2009, DeLeo2011}: the MI state is signaled by a suppressed double occupancy while the CDW order is formed by atoms on alternating doubly occupied sites.

%P8:
\begin{figure}[bt]
    \includegraphics{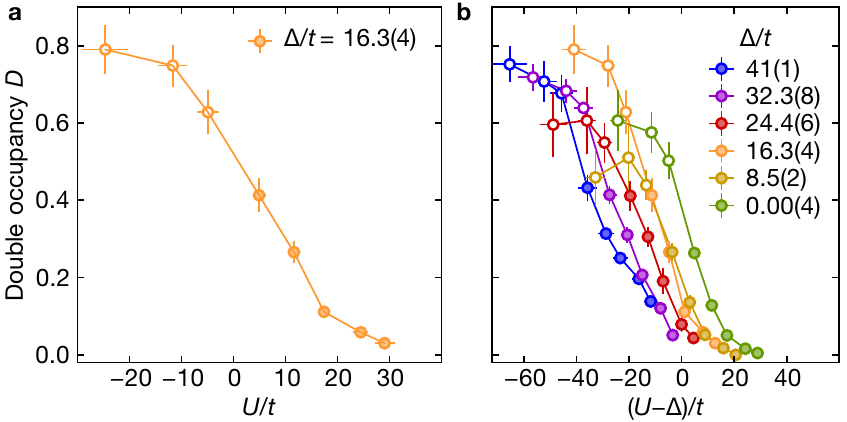}
    \caption{Double occupancy measurement
    {\bf (a)} The measured double occupancy $D$ as a function of the on-site interaction $U$ for a fixed energy-offset $\Delta=16.3(4)t$. 
    {\bf (b)} For different values of $\Delta$ (different colors) we obtain the double occupancy for a range of interactions $U=[-24.6(13),29.1(7)]t$.
    Hollow (full) circles represent attractive (repulsive) interactions. 
    Vertical error bars show the standard deviation of 5 measurements and horizontal error bars the uncertainty on our lattice parameters.    
	}\label{fig2}
\end{figure}
In the experiment we set an energy-offset $\Delta$ and measure $D$ for different attractive and repulsive interactions $U=[-24.6(13) , +29.1(7)]t$.
Fig. \ref{fig2}a shows $D$ as a function of $U$ at constant $\Delta=16.3(4)t$.   
For strong attractive interactions we observe a large fraction of doubly occupied sites, which continuously decreases as $U$ is increased. 
When tuning from attractive to weak repulsive interactions $(\Delta \gg U)$, we still observe a large $D$ as expected for the CDW. 
For strong repulsive interactions $(U \gg \Delta)$ the measured double occupancy vanishes, the density pattern no longer reflects the broken inversion symmetry of the lattice, confirming the suppression of the CDW ordering. 
Fig. \ref{fig2}b shows $D$ as a function of the energy scale $U-\Delta$, which is the energy difference of a doubly occupied site neighbouring an empty site compared to two singly occupied sites in the atomic limit. 
For the largest negative value of $U-\Delta$ we observe the highest $D$ for all $\Delta$. 
For positive values of $U-\Delta$ the double occupancy continuously decreases and vanishes for the largest positive $U-\Delta$, consistent with a MI state. 
In contrast, for the intermediate regime the measured $D$ depends on the individual values of $U$ and $\Delta$, as now the finite temperature and chemical potential itself play an important role and a detailed analysis would be required for a quantitative understanding, however we can qualitatively compare the dependence of $D$ to an atomic limit calculation \cite{supplementary}.

%P9:
\begin{figure}[t!]
    \includegraphics{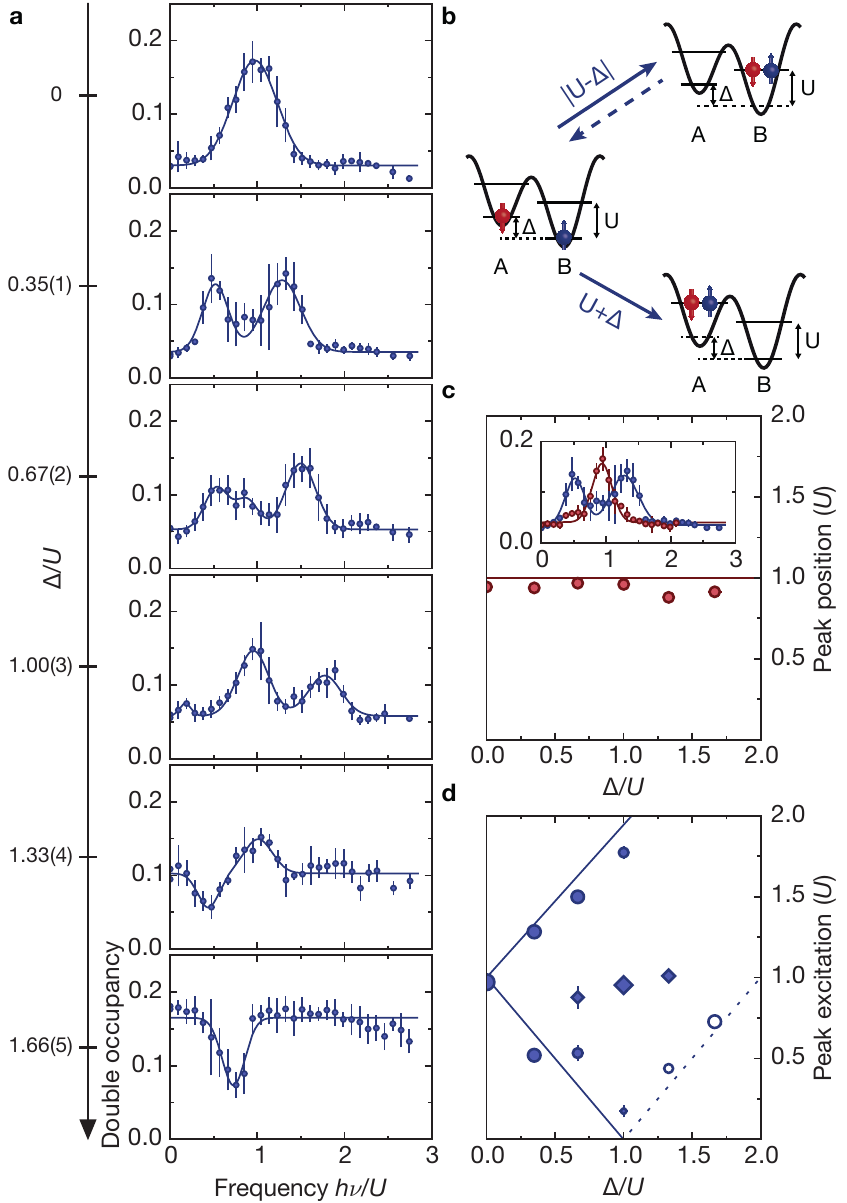}
    \caption{Modulation spectroscopy measurement
    {\bf (a)} Excitation spectra observed by measuring the double occupancy $D$ from amplitude modulation spectroscopy of the lattice beam in y-direction for different energy-offsets $\Delta$ at repulsive on-site interaction $U=24.4(5)t$. 
    Solid lines are multiple Gaussian fits to the modulation spectra. 
    {\bf (b)} Schematics for the relevant energy scales $|U-\Delta|$ and $U+\Delta$ as a response to the lattice modulation. 
    {\bf (c)} Modulation spectroscopy of the lattice beam in z-direction. 
    The measured excitation frequencies are shown as a function of $\Delta$ and compared to the value of $U=24.4(5)t$ (horizontal line).
    The inset shows the spatially dependent excitation spectrum. 
    {\bf (d)} Comparison of the measured excitation resonances (points) with the values of $|U-\Delta|$, $U+\Delta$ (lines). 
    The area of the marker indicates the strength of the response (peak height) to the lattice modulation. 
    Full (empty) circles represent a positive (negative) response in double occupancy.  
    Error bars as in Fig. \ref{fig2}, vertical error bars in {\bf (c)},{\bf (d)} show the fit error for the peak position.
	}\label{fig3}
\end{figure}
A characteristic feature of the MI and band insulating CDW state is a gapped excitation spectrum, which we probe using amplitude-modulation spectroscopy \cite{Joerdens2008, Kollath2006}.
We sinusoidally modulate the intensity of the lattice beam in y-direction by $\pm10\%$ for $40 \, \mathrm{ms}$.
Since the honeycomb lattice is created from several beams interfering in the xy-plane \cite{Tarruell2012}, this leads to a modulation in tunnel coupling $t_y$ of 20\% and $t_x$ of 8\%, as well as a modulation of $U$ by 4\% and $\Delta$ by up to 6\%. 
The interlayer tunnelling $t_z$ is not affected meaning that excitations only occur in the honeycomb plane. 
We set $U=24.4(5)t$ and measure $D$ after the modulation for frequencies up to $\nu = 11.6 \, \mathrm{kHz}$ ($\approx 67 \, t$). 
All measurements are performed in the quadratic-response regime \cite{Greif2011}.

%P10: 
Fig. \ref{fig3}a shows the measured spectra for different values of $\Delta$. 
The MI state exhibits a gapped excitation spectrum, which is directly related to a particle-hole excitation with a gap of size $U$ \cite{Joerdens2008, Greif2011, Uehlinger2013}.
In the limit of $\Delta=0$ we detect this gap as a peak in the excitation spectrum at $\nu=U/h$. 
With increasing $\Delta$ the single excitation peak splits into two peaks corresponding to different excitation energies \footnote{Related excitations have been observed with bosons in tilted optical lattices \cite{Sias2008, Ma2011, Meinert2013} and for two fermions in a single double-well \cite{Murmann2015}}.  
The nature of the excitations can be understood as follows:  
The transfer of one particle costs approximately an energy of $U-\Delta$ if a double occupancy is created on a {\bf B} site and $U+\Delta$ if it is created on an {\bf A} site (see Fig. \ref{fig3}b). 
The excitation of additional double occupancies shows that atoms were initially populating both sub-lattices, as expected in the MI regime. 
For small $\Delta/U$ the system shows a clearly identifiable charge-gap, which vanishes if $U \sim \Delta$.
For large $\Delta$ the charge gap reappears, and a minimum in the spectra reveals the breaking of double occupancies as a response to amplitude modulation. 
This is in agreement with the expected band insulating CDW, where double occupancies are on the {\bf B} sub-lattice and {\bf A} sites are empty.

%P11:
The situation changes for amplitude modulation of the z lattice beam intensity by $\pm10 \%$.
In this case excitations are created along links perpendicular to the honeycomb plane. 
Since the honeycomb lattice is replicated along the z-axis, we observe a single peak at $\nu = U/h$, independent of the energy-offset $\Delta$ (see Fig. \ref{fig3}c).  
The inset of Fig. \ref{fig3}c shows the direction dependent modulation spectrum for $\Delta=8.5(2)t$, which allows us to independently determine the energy scales of the system in different spatial directions.

%P12:
We extract the excitation energies by fitting multiple Gaussian curves to our experimental data and compare our results with the values of $|U-\Delta|$, $U+\Delta$ and $U$ in Fig. \ref{fig3}d.
We observe a vanishing peak at $U+\Delta$ for the largest $\Delta$.  
This is expected as there are fewer and fewer atoms on {\bf A} sub-lattice in the system for an increasing energy-offset.
Our measurements are in good agreement with a picture based on nearest-neighbour dynamics.

%P13:
\begin{figure}[bt]
    \includegraphics{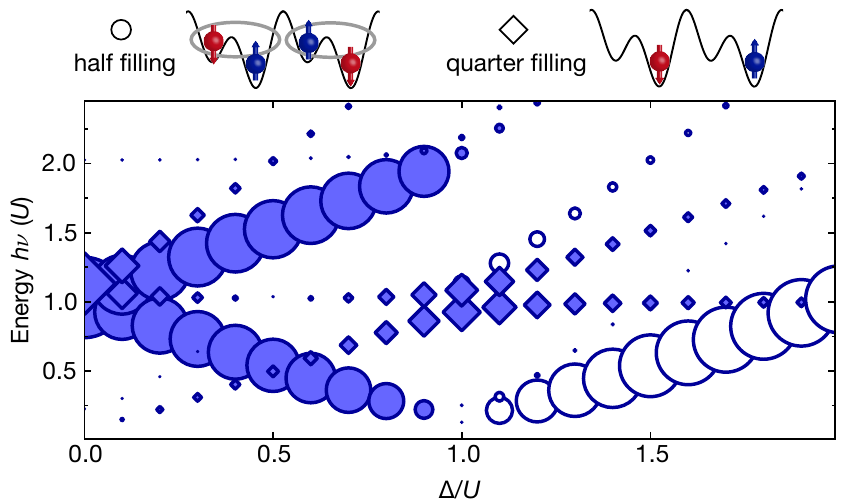}
    \caption{Theoretical result for the kinetic energy response function $\chi(\nu)$ of the double occupancy on a modulated four site model as a function of $\Delta$ at constant $U=25t$. 
    Circular (diamond) data points represent the response for the half filled (quarter filled) case. 
    The area of the marker shows the relative size of the calculated response, whereas full (empty) data points have a positive (negative) response signal. 
	}\label{fig4}
\end{figure}
However, we observe additional peaks at $\nu \approx U /h$ if $U \sim \Delta$, which can not be understood in a two-site model.
To rule out any higher-order contribution, we verified that the response signal has a quadratic dependence on the modulation parameters, as expected for quadratic response \cite{Greif2011}.
This additional peak was also observed in a purely 2D ionic Hubbard model ($t_z=h\times2$ Hz), thus ruling out a contribution of excitations along the third direction \cite{supplementary}.

%P14:
To interpret the nature of the response at $h \nu \approx U$ we calculate the kinetic energy response function 
\begin{equation}
\chi(\nu) = \sum_{m} \langle m |\delta D | m \rangle |\langle m |K| 0\rangle|^{2} \delta(h\nu-\epsilon_{m0}),
\end{equation}
where the sum runs over all many-body states $m$, $\delta D=D-\langle 0| D|0\rangle$ is the induced change in double occupancy, $K = \sum_{\langle ij \rangle,\sigma}\hat{c}_{i\sigma}^{\dagger}\hat{c}_{j\sigma}$ and $\epsilon_{m0}$ denotes the excitation energy measured above the ground state $| 0\rangle$. 
We evaluate $\chi(\nu)$ in exact diagonalization of a cluster of four sites for varying filling fractions.

%P15:
The result shown in  Fig.~\ref{fig4} for $U/t=25$ clearly indicates that the peak at $h \nu \approx U$ around $U=\Delta$ originates from regions of the lattice where the filling deviates from one particle per site \footnote{Due to the harmonic confinement our system exhibits a large fraction of atoms away from half-filling, for more details see \cite{supplementary}}. 
In particular, for a configuration with two particles on four sites, the ground state at $U=\Delta$ is a configuration with negligible double occupancy and only the lower sub-lattice sites are filled. The lattice modulation at $h \nu \approx U$ then moves one particle to an energetically costly site. 
For $U=\Delta$, this configuration is resonantly coupled to a state where both particles are on the same, low-energy site. 
Hence, this process leads to an increase in the measured double occupancy. 
The analysis of such a four-site cluster qualitatively agrees with the observed signal at energy $U$ in the intermediate ($U\approx \Delta$) regime.

%P16:
In conclusion, we have realized and studied the ionic Hubbard model with ultracold fermions in an optical honeycomb lattice. 
Our observations show that increasing interactions suppress the CDW order and restore inversion symmetry of the density distribution.
Additionally, we probed correlations beyond nearest-neighbour, which had not been accessible so far \cite{Greif2013}. 
Future work can address open questions concerning the nature of the intermediate regime between the two insulating phases, which is theoretically debated and should depend on the dimensionality of the system \cite{Garg2006, Paris2007}.  
Furthermore, we can extend our studies of the ionic Hubbard model to include topological phases by introducing complex next nearest neighbour tunnelling \cite{Jotzu2014, Prychynenko2014, Zheng2015}.

\begin{acknowledgments}
We thank Ulf Bissbort, Frederik G\"org, Diana Prychynenko, Vijay Shenoy, Leticia Tarruell, Evert van Nieuwenburg and Lei Wang for insightful discussions. 
We acknowledge support from the SNF, NCCR-QSIT, QUIC (EU H2020-FETPROACT-2014) and SQMS (EU ERC advanced grant). 
R.D. acknowledges support from ETH Zurich Postodoctoral Program and Marie Curie Actions for People COFUND program.
\end{acknowledgments}

\clearpage

\section{Supplemental Material}  
  
\subsection{Preparation and optical honeycomb lattice}

To simulate the ionic Hubbard model we create a quantum degenerate cloud of $^{40}\, \mathrm{K}$ by using a balanced spin mixture of the $\vert F, m_{F}\rangle = \vert 9/2,-9/2 \rangle $ and $\vert 9/2,-7/2 \rangle$ Zeeman states, which is evaporatively cooled in an optical dipole trap. 
Depending on the desired interaction strength we either use the Feshbach resonance of the $m_{F}=-9/2,-7/2$ mixture to access an interaction range of $U = [-24.6(13),4.91(9)]t$ or a $m_{F}=-9/2,-5/2$ mixture to reach strongly repulsive interaction strengths $U = [11.7(2),29.1(7)]t$.
The fermions, with temperatures between 16(2)\% and 13(2)\% of the Fermi-temperature, are then loaded into a three-dimensional optical lattice within $200 \, \mathrm{ms}$. The honeycomb lattice with staggered energy-offset $\Delta$ is created by interfering laser beams at a wavelength $\lambda=1064 \, \mathrm{nm}$ that give rise to the following potential \cite{Tarruell2012}: 
\begin{eqnarray} V(x,y,z) & = & -V_{\overline{X}}\cos^2(k
x+\theta/2)-V_{X} \cos^2(k
x)\nonumber\\
&&-V_{Y} \cos^2(k y) -V_{\widetilde{Z}} \cos^2(k z) \nonumber\\
&&-2\alpha \sqrt{V_{X}V_{Y}}\cos(k x)\cos(k
y)\cos\varphi , \label{eqlattice}
\end{eqnarray}
where $V_{\overline{X},X,Y,\widetilde{Z}}$ are the single beam lattice depths in each spatial direction, $k=2\pi/\lambda$, and the visibility $\alpha=0.90(5)$. 
We interferometrically stabilize the phase $\varphi=0.00(3)\pi$ and control the energy-offset $\Delta$ between the {\bf A} and {\bf B} sub-lattice by varying the value of $\theta$ around $\pi$.
This is achieved by changing the frequency detuning between the $\overline{X}$ and the $X$ (which has the same frequency as $Y$) beam.
In the case of $\Delta=0$ the lattice depths are set to $V_{\overline{X},X,Y,\widetilde{Z}}=[14.0(4),0.79(2),6.45(20),7.0(2)]E_{R}$ to prepare isotropic tunnelling bonds in the honeycomb lattice ($t/h =174(12)\, \mathrm{Hz}$), where $E_{R}$ is the recoil energy. 
When breaking inversion symmetry ($\Delta \neq 0$) we adjust the final lattice depths in order to keep $t$ on all lattice bonds constant.
Owing to the harmonic confinement of the lattice beams and the remaining dipole trap our system has harmonic trapping frequencies of $\nu_{x,y,z}=[55.6(7),106(1),57(1)]\, \mathrm{Hz}$. 
The lattice depths are independently calibrated using Raman–Nath diffraction on a $^{87}\mathrm{Rb}$ Bose–Einstein condensate. 
For the measurement of the double occupancy $D$, both the independently determined offset in $D$ of 2.2(3)\% due to an imperfect initial spin mixture as well as the calibrated detection efficiency of 89(2)\% for double occupancies are taken into account \cite{Greif2013}.

\subsection{Noise correlations}

The anti-correlations in the fluctuations of the momentum distribution stem from the fermionic nature of the particles. To illustrate this, let us consider the simple case of two identical fermions, occupying the lowest band of a one-dimensional optical lattice of periodicity $\lambda/2$. 
Their wavefunctions can be expressed as Bloch waves, with a different quasi-momentum $q$ for each particle, such that they obey the Pauli principle.  
If the momentum of the first particle is measured to be $q_1$, then the second cannot be detected in any of the momenta $q_1-n\hbar 4\pi/\lambda$ with $n$ an integer.

In the experiment, the momentum distribution $\tilde{n}(q,\tau=0)$ is accessed by suddenly releasing the atomic cloud from its confinement. 
After an expansion time sufficient to neglect the initial size of the cloud, the momentum distribution has been converted to the spatial atomic density $n(x,\tau)$ which is then directly measured by taking an absorption image.  
These two quantities are related by
\begin{equation}
n(x,\tau) = \tilde{n}(q=\frac{m\,x}{\hbar \tau},\tau=0)
\end{equation}
In the following we use $n$ to designate the atomic density in momentum space. In general, we are interested in the probability to detect two particles at momenta $q_1$ and $q_2$ simultaneously, which is given by
\begin{align}
P(q_1,q_2)&=\langle n(q_1) n(q_2) \rangle - \langle n(q_1) \rangle \langle n(q_2) \rangle \nonumber \\
&= \sum_{\alpha,\beta} \langle \hat{\Psi}^\dagger_\alpha(q_1) \hat{\Psi}_\alpha(q_1) \hat{\Psi}^\dagger_\beta(q_2) \hat{\Psi}_\beta(q_2) \rangle \nonumber \\ 
&- \langle \hat{\Psi}^\dagger_\alpha(q_1) \hat{\Psi}_\alpha(q_1) \rangle \langle\hat{\Psi}^\dagger_\beta(q_2) \hat{\Psi}_\beta(q_2) \rangle ,
\end{align}
where $\hat{\Psi}^\dagger_\alpha(q)$ creates and $\hat{\Psi}_\alpha(q)$ annihilates a particle with internal state $\alpha$ at momentum $q$. 
Representing the operators by their underlying Bloch-wave structure we have
\begin{align} 
P(q_1,q_2) &= |W(q_1)|^2 |W(q_2)|^2 \sum_{\alpha, \beta} \sum_{k,l,m,n}  e^{i\,\lambda/2(q_1(k-m)+q_2(l-n))} \nonumber \\
& \times \left( -\langle \hat{b}^\dagger_{\alpha,k}\,\hat{b}^\dagger_{\beta,l}\,\hat{b}_{\alpha,m}\,\hat{b}_{\beta,n}\rangle-\langle \hat{b}^\dagger_{\alpha,k}\,\hat{b}_{\alpha,m}\rangle \langle \hat{b}^\dagger_{\beta,l}\,\hat{b}_{\beta,n}\rangle \right) \nonumber \\
&+\delta(q_1-q_2)\sum_{\alpha}\langle \hat{\Psi}^\dagger_\alpha(q_1) \hat{\Psi}_\alpha(q_2) \rangle , 
\end{align}
where $\hat{b}^\dagger_{\alpha,k}$ ($\hat{b}_{\alpha,k}$) creates (annihilates) a particle at the site $k$ with an internal state $\alpha$. 
Here, $W(q)$ is the slowly varying envelope of the Bloch wave. 
The last term concerns the correlation of an atom with itself, which we are not considering here. 
To calculate the four-operator expectation value, we assume that the atomic distribution is well described by Fock states ($\langle \hat{b}^\dagger_{\alpha,k} \hat{b}_{\beta,l} \rangle =\delta_{k,l} \delta_{\alpha,\beta} n_k $), where $n_k$ is the number of particles on site $k$. 
Thus, we have
\begin{align}
P(q_1,q_2) &= -|W(q_1)|^2 |W(q_2)|^2 \nonumber \\
& \times \sum_{\alpha} \sum_{k,l} n_{\alpha,k} n_{\alpha,l} e^{i\,\lambda/2(k-l)(q_1-q_2)} \\
P(q_0,d)&= -|W(q_0+d/2)|^2 |W(q_0-d/2)|^2 \nonumber \\
& \times \sum_{\alpha} \sum_{k,l} n_{\alpha,k} n_{\alpha,l} e^{i\,\lambda/2(k-l)d} , 
\end{align}
where we have introduced the center of mass $q_0=(q_1+q_2)/2$ and the relative position $d=q_1-q_2$. 
The slowly varying dependence in $q_0$ can be rewritten in terms of the momentum distribution
\begin{equation}
\langle n(q_1) \rangle \langle n(q_2) \rangle = |W(q_0+d/2)|^2 |W(q_0-d/2)|^2 N^2
\end{equation}
with $N$ the total atom number. 
Thus, the correlations in momentum are fully characterized by
\begin{align}
C(d)&=\frac{\int \mathrm{d}q_0 P(q_0,d)}{\int \mathrm{d}q_0 \langle n(q_0+d/2) \rangle \langle n(q_0-d/2) \rangle} \nonumber \\
&= -\sum_{\alpha} \sum_{k,l} \frac{n_{\alpha,k} n_{\alpha,l}}{N^2} e^{i\,\lambda/2(k-l)d}
\end{align}

\begin{figure}
    \includegraphics{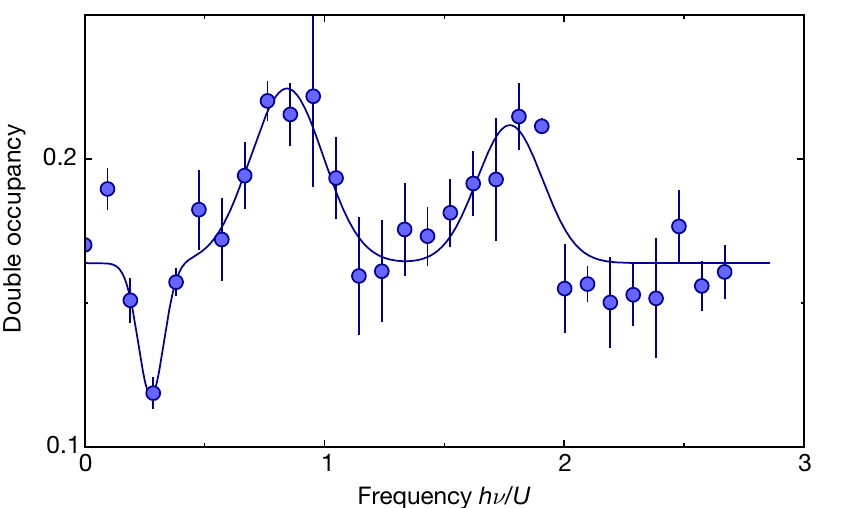}
    \caption{Modulation spectroscopy measurement for the two-dimensional ionic honeycomb potential.
    Excitation spectra observed by measuring the double occupancy $D$ after sinusoidal modulation of the lattice depth $V_y$ for an energy-offset $\Delta=24.3(6)t$ at constant repulsive on-site interaction $U=24.1(4)t$, thereby realizing the intermediate regime $|U-\Delta| \sim t$.
    Although the tunnelling perpendicular to the honeycomb planes is tuned below $2\, \mathrm{Hz}$ we still observe the response at energy $\approx U$.
    Error bars show the standard deviation of at least 3 measurements.   
   	}\label{sfig1}
\end{figure}

We now show that this quantity is not only sensitive to the periodicity imposed by the lattice, but can also reveal underlying order in the density distribution. 
Consider the case where, for each internal state, the density takes the value $n_{\bf A}/M$ on even-numbered sites and $n_{\bf B}/M$ on odd numbered sites, with $M$ the number of internal states. The correlation signal then takes on the form
\begin{equation}
C(d)=-\frac{n_{\bf A}^2+n_{\bf B}^2+2\,n_{\bf A}\,n_{\bf B}\cos(\lambda/2\,d)}{M\,N^2}\sum_{k,l}e^{i\,\,\lambda(k-l)d}
\end{equation}
The sum is equal to $N^2$ if $d=m\,2\pi/\lambda$ with $m$ an integer and zero otherwise, and the correlation signal is then simply
\begin{equation}
C(m\,2\pi/\lambda)=-\frac{(n_{\bf A}+(-1)^m\,n_{\bf B})^2}{M}
\end{equation}
Thus, anti-correlations always appear at momenta $2m\times 2\pi/\lambda$, corresponding to the reciprocal lattice vector, while anti-correlations at momenta $(2m+1)\times 2\pi/\lambda$ signal a staggering of the atomic density between the even- and odd-numbered sites. While this derivation was carried out for a one-dimensional lattice, it can readily be generalized to higher dimensions, provided the full information on the momentum density can be accessed.

In the experiment, we measure the momentum distribution of the absorption images following a three step detection protocol.  
After preparing the system in a shallow honeycomb lattice with a given $U$ and $\Delta$, the lattice depth is suddenly increased in $1 \, \mathrm{ms}$, which prevents any further evolution of the atomic density distribution.  
Subsequently, the lattice geometry is converted to a simple cubic lattice within $1 \, \mathrm{ms}$. 
Measuring the density distribution in the honeycomb lattice would lead to additional peaks at $\mathbf{m}=(\pm 1,\pm 1)$ due to the displacement of the lattice sites with respect to a square lattice.  
We can estimate the strength of these additional peaks with a simple model for a hexagonal lattice with $\Delta=0$ by placing Gaussian wave packets at the position of each lattice site of the real potential. 
By calculating the Fourier transform of this system we find that the strength of the \textbf{$\mathbf{m}=(\pm 1,\pm 1)$} peaks is a factor of 6 smaller than the minima of the correlator at position $\mathbf{m}=(0,\pm 2)$ and $\mathbf{m}=(\pm 2,0)$. 
Therefore ramping to a simple cubic configuration ensures our observable probes correlations of the underlying density order rather than a specific lattice structure. 
Finally, the strength of the interactions is reduced within $50 \, \mathrm{ms}$ using the Feshbach resonance, the atoms are released from the lattice and left to expand ballistically for $10\, \mathrm{ms}$ after which we measure the density distribution by absorption imaging.

From this measurement, we compute the quantity $C$, which is shown in Fig. 1. 
As our imaging technique integrates the density along the line of sight, we do not have access to the full information, but rather to the column density. 
Thus, the derivation presented above should be generalized to two dimensions, while the occupancy along the third direction can be treated as an internal degree of freedom.
Accordingly, the depth of the anti-correlation minima for a two-component Fermi gas will be divided by $2\,N_z$, where $N_z$ is the typical number of sites populated along the integrated direction.
To achieve optimal signal-to-noise, we only consider atomic densities above 20 \% of the maximum density. 
Furthermore, we remove short-range correlations induced by the readout noise of the CCD chip of the camera by convoluting the density distribution with a Gaussian of width $\Delta q=k/25$. 
Finally, we take advantage of the reflection symmetry of the momentum distribution, and average together $C(d_x,d_y)$ with $C(d_x,-d_y)$. Note that, by definition, $C(d_x,d_y)=C(-d_x,-d_y)$.

\subsection{Modulation spectroscopy}

The tunnelling $t_{z}$ between sites of adjacent layers can be controlled via the lattice depth $V_{\tilde{Z}}$. 
To realize the ionic Hubbard model in two dimensions we suppress the tunnelling $t_{z}$ below $2\, \mathrm{Hz}$ by setting the lattice depth along the z-direction to $V_{\tilde{Z}}=30(1)E_{R}$ thereby decoupling the honeycomb planes.
For the modulation spectroscopy measurement we follow the same procedure as described in the main text and sinusoidally modulate the amplitude of the lattice depth in y-direction by $\pm 10\%$.
As a result we exclude possible contributions to the response signal, which may result from a residual coupling to the orthogonal direction.
This ensures that the linear response measurement only probes energy scales that are realized within the xy-honeycomb plane. 
Fig. \ref{sfig1} shows the excitation spectrum for the two-dimensional case. 
Even with suppressed tunnelling $t_{z}$ we observe a clear peak for modulation frequencies $h\nu = U$. 
As a result we can exclude that our response signal is resulting from an imperfect orthogonality of the lattice beams.

\subsection{Atomic limit calculation and local density approximation}

\begin{figure}
    \includegraphics{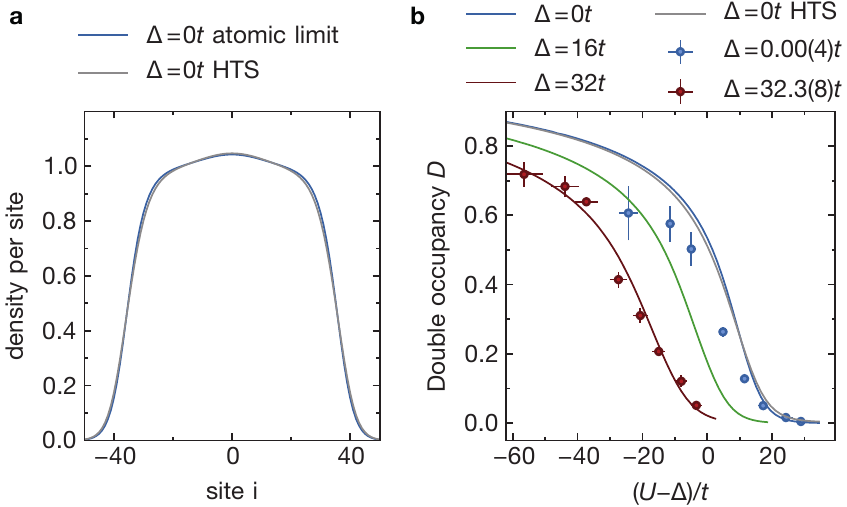}
    \caption{Theory comparison of the density per site in the trapped system and calculation of the double occupancy.
    For the calculation we use an entropy per particle of $1.5 k_B$ (as measured before loading the atomic gas into the lattice) and a total atom number of 190000. 
 {\bf (a)} Density per site calculated for $U=25t$ and $\Delta=0$ using a local density approximation to include the harmonic trap.  
  The high-temperature series (HTS) expansion up to second order of the grand canonical partition function is shown in grey. 
  The atomic limit calculation (high-temperature series expansion in 0th order) is plotted in blue. 
    {\bf (b)} For different values of $\Delta$ (different colors) we calculate the double occupancy in the atomic limit using a local density approximation.
    We can compare the atomic limit calculation (blue) with a high-temperature series expansion up to second order (grey) for $\Delta=0$.
    Points show the measured double occupancy as plotted in Fig. 2b.   
   	}\label{sfig2}
\end{figure}

We can qualitatively estimate the density per site, temperature and double occupancy of our system from an atomic limit calculation (high-temperature series expansion of the partition function in 0th order) including the harmonic trap via a local density approximation (i.e. locally varying the chemical potential). 
While this is not an exact theory capturing all details and not suitable for a precise experiment-theory comparison, this calculation gives a qualitative estimate for the typical system parameters.
In the case of $\Delta=0$ we can directly compare the obtained results for the density of atoms per site with a high-temperature series expansion up to second order \cite{Uehlinger2013} of the grand canonical partition function, see Fig. \ref{sfig2}a. 
As expected, when comparing the results for the density of atoms per site we see a small deviation between the two theories. 

Using the atomic limit calculation with the local density approximation we can estimate the fraction of half-filling and quarter-filling in our trapped system. 
Approximately a fraction of 46\% of the atoms are within 10\% of half-filling for values $U=25t$ and $\Delta=0t$.
The nature of the system totally changes for $U = \Delta =25t$, as a large fraction of the atoms (45\%) consists of a quarter filled region. 
Deep in the two limiting cases ($U \gg \Delta$ and $\Delta \gg U$), we expect the approximation to provide better results.
Given an entropy per particle of $1.5 k_B$ (as measured before loading the atomic gas into the lattice), and assuming adiabatic loading into the lattice, we find for the calculated temperature $T$ values $3.3t$  $(5.5t)$, for parameters of $U=30t$ and $\Delta=0$ $(U=10t$ and $\Delta=40t)$.
As $\Delta$ and $U$ are much larger than $t$, this confirms that the temperature is well below the charge gap for the MI and CDW states.

For a qualitative comparison of our measurements of the double occupancy in Fig.2 we can calculate the fraction of double occupied sites $D$ for various values of $U$ and $\Delta$ using our simplified model at constant entropy per particle of $1.5 k_B$. 
As can be seen in Fig. \ref{sfig2}b our results qualitatively confirm our observations that for increasing $U$ the double occupancy is reduced well before reaching the point $U=\Delta$.
This calculation shows that the major contribution of the shift is resulting from a change in the chemical potential and temperature for different values of $U$ and $\Delta$. 
For $\Delta=0$ we can estimate the importance of the tunnelling $t$ by directly comparing the double occupancy obtained from the atomic limit calculation with a high-temperature series expansion up to second order.

\makeatletter
\setcounter{section}{0}
\setcounter{subsection}{0}
\setcounter{figure}{0}
\setcounter{equation}{0}
\renewcommand{\bibnumfmt}[1]{[S#1]}
\renewcommand{\citenumfont}[1]{S#1}
\renewcommand{\thefigure}{S\@arabic\c@figure}
\renewcommand{\theequation}{S\@arabic\c@equation}
\makeatother

\end{document}